# Factors Affecting Acceptance of Web-Based Training System: Using Extended UTAUT and Structural Equation Modeling

Thamer A. Alrawashdeh, Mohammad I. Muhairat and Sokyna M. Alqatawnah

Department of software Engineering, Alzaytoonah University of Jordan, Amman, Jordan

Thamer.A@zuj.edu.jo

Drmohoirat@zuj.edu.jo

S.Alqatawnah@zuj.edu.jo

## ABSTRACT

*Advancement in information system leads organizations to apply e-learning system to train their employees in order to enhance its performance. In this respect, applying web based training will enable the organization to train their employees quickly, efficiently and effectively anywhere at any time. This research aims to extend Unified Theory of Acceptance and Use Technology (UTAUT) using some factors such flexibility of web based training system, system interactivity and system enjoyment, in order to explain the employees' intention to use web based training system. A total of 290 employees have participated in this study. The findings of the study revealed that performance expectancy, facilitating conditions, social influence and system flexibility have direct effect on the employees' intention to use web based training system, while effort expectancy, system enjoyment and system interactivity have indirect effect on employees' intention to use the system.*

## 1. INTRODUCTION

Nowadays, with the development of the Word Wide Web, e-learning system provides many benefits to individuals and organizations. Additionally, it enables the employees to access the training materials from any way at any time which have overcomes the many challenges with the traditional training methods. Also, such system offers an enjoyable training environment due to the presentation of the materials in various forms (e.g. video, audio, animation and etc.). Furthermore, with the increasing demand to improve the employees' skills and knowledge reflecting on their work performance and their productivity, web based training system enables the organizations to offer the training for their employees without any adversely effect on work performance. Additionally, web based training reduces the training cost and time and support the customers [12].

However, not many studies have been conducted to the acceptance of such system [25], [7] till now. Therefore, this study investigates the acceptance of web based training.

In the meantime, research on acceptance of e-learning system by universities' students and organizations' employees has generated interest of a lot of information system researchers. They have identified many constructs that influence people intention to use e-learning system [12], [17], [14], [26]. These researchers had used many models and theories to explain the acceptance of information technology. In this respect, the modern model had been used to describe such acceptance is Unified Theory of Acceptance and Use Technology (UTAUT) [24]. As far as, there is no trail to extend this theory to include other successful factors for e-learning system acceptance. This study makes an effort to extend the original UTAUT to include three critical success factors in the e-learning context including, system flexibility, system enjoyment and system interactivity [12], [17], [1], [19], [7] (see figure 1).

## 2.0 THEORETICAL FRAMEWORK AND HYPOTHESES

### 2.1 Unified Theory of Acceptance and Use Technology (UTAUT)

The acceptance of web based training system may be treated as information system acceptance. As previously mentioned the modern theory in this field is UTAUT [24]. This theory could predict the acceptance of an information system in approximately 70% of the cases. Comparing with TAM, it could only predict the acceptance of an information system in approximately 40% of the cases. On the other hand, the validity of UTAUT in the information system context needs further testing [14].

Therefore in this study the extended UTAUT with some of information system successful factors that mentioned below in this section is going to be tested. Thus, the following hypotheses have been proposed for this study.

H1. Performance expectancy will have direct effect on the employees' intention to use web based training system.

H2. Effort expectancy will have direct effect on the employees' intention to use web based training system.

H3. Social influence will have direct effect on the employees' intention to use web based training system.

H4. Facilitating conditions will have direct effect on the employees' intention to use web based training system.

### 2.2 System Flexibility

Many scholars introduced the perceived flexibility as one of the critical factors to understand user's behavioral acceptance of e-learning system [12], [17], [11], [15]. Flexibility of e-learning system was defined as the degree to which individual believes that he/she can access the system from anywhere at any time [12]. Adapting this construct to examine the acceptance of web based training system by public sector's employees suggests that they will accept web based training system if they believe that they can access the system from anywhere at any time. Hsia and Tseng [12], Sahin and Shelley [17], Nanayakkara [16] and Lim et al. [13] argued that perceived flexibility of e-learning system positively influence user's intention to use e-learning system. Therefore, the following hypothesis is proposed.

H5. System flexibility has a positive effect on employees' intention to use web based training system.

### 2.3 System Interactivity

Although, few studies have paid attention to this factor, Abbad et al. [1] suggested that system interactivity has indirect impact on the user's intention to use e-learning system through perceived usefulness and perceived ease of use. Additionally, Davis [10] found that perceived usefulness and perceived ease of use fully mediates effect of system's characteristics on user's intention to use the e-mail technology. Consequently, because many scholars agree that perceived performance expectancy and perceived effort expectancy similar to perceived usefulness and perceived ease of use [24], [26], [14] the following hypotheses are proposed.

H6. System interactivity has a positive impact on perceived performance expectancy.

H7. System interactivity has a positive impact on perceived effort expectancy.

### 2.4 System enjoyment

In the effect of perceived system enjoyment, many studies indicated that perceived system enjoyment has direct effect on user's intention to use e-learning system and indirect effect on the user's intention through perceived ease of use and perceived usefulness [19], [7], [9]. Thus, the following hypotheses are proposed.

H8. System enjoyment has a direct impact on perceived performance expectancy.

H9. System enjoyment has a direct impact on perceived effort expectancy.

H10. System enjoyment has a direct impact on employees' intention to use web based training.

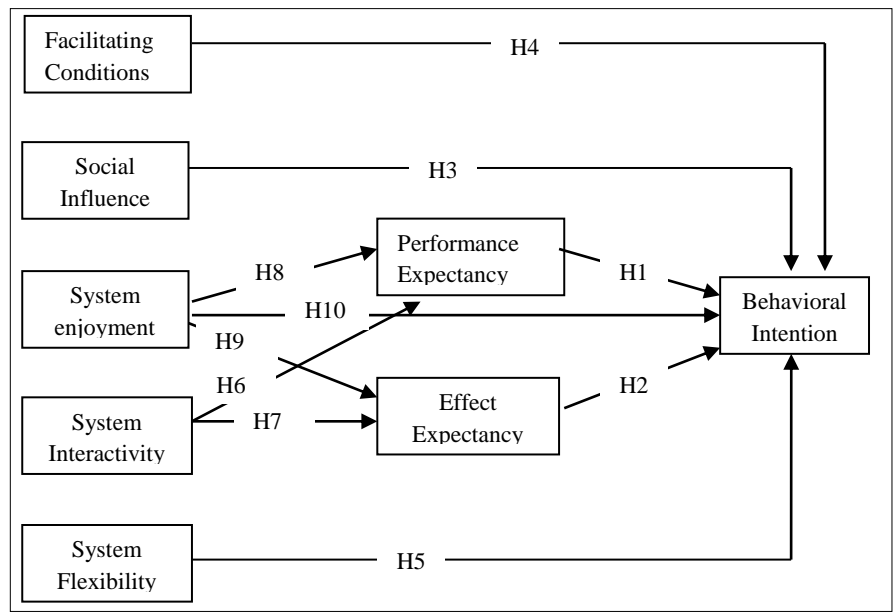

Figure 1. Theoretical research framework

## 3.0 RESEACH METHODOLOGY

### 3.1 Data collection method

A questionnaire has been designed and used to collect a data. This research is going to measure eight constructs and the questionnaire was divided into nine sections. The first section includes information regarding the characteristics of respondents (e.g. age, gender, having personal computer, having internet access, having experience with e-learning system), while each one of other sections includes questions that measure each of this research model constructs. The total number of questionnaire's items is 43. Each item is measured using 7point-likert scale. All such items have been adapted from [1], [17], [24].

### 3.2 Sampling and Content Validation

The validity is concerned with reducing the possibility of getting incorrect answers during the data collection period [18]. In this research, content validity was carried out through questionnaire pre-test process [27], while, the questionnaire was modified based on the comments which were received from ten employees who responded to the questionnaire before it was distributed to the sample of study.

In total, five hundred (500) questionnaires had been distributed to the public sector's employees in Jordan. Eventually, only two hundred and ninety employees at a response rate of 58% had successfully completed and returned the questionnaire. Lately, the Structural Equation Model (SEM) approach and AMOS software was used to analyze the data.

## 4.0 DATA ANALYSIS AND RESULTS

### 4.1 Measure Reliability and construct Validity

AMOS 16.0 statistical software (structural equation model) was used to evaluate the construct validity and the reliability of the measurement. Interestingly, the following equation was used to measure the composite reliability $(\sum \text{factor loading})2 / (\sum \text{factor loading})2 + \sum \text{measurement error}$. As well as, the average variance extracted (AVE) was measured to examine the convergent validity using the following equation AVE= $\sum (\text{factor loading})2/\sum(\text{factor loading})2 + \sum \text{measurement error}$. The results of composite reliability and convergent validity tests provided evidence of the validity of the measurement's items, since the reliability and validity of all the constructs have exceeded the recommended level of 0.7 (see table 2).

Table 2. Instrument reliability and validity

| Constructs | Items | Loadings | Composite reliability >= 0.7 | Convergent Validity >= 0.7 |
|---|---|---|---|---|
| Performance Expectancy | PE 2 | .956 | 0.97 | 0.93 |
|  | PE 3 | .965 |  |  |
|  | PE 4 | .963 |  |  |
| Effort Expectancy | EE 5 | .972 | 0.96 | 0.94 |
|  | EE 6 | .958 |  |  |
| System Interactivity | SIN 1 | .921 | 0.96 | 0.93 |
|  | SIN 2 | .928 |  |  |
|  | SIN 3 | .951 |  |  |
| System Enjoyable | SE 1 | .960 | 0.95 | 0.95 |
|  | SE 2 | .947 |  |  |
| System Flexibility | SF 1 | .974 | 0.96 | 0.95 |
|  | SF 2 | .964 |  |  |
|  | SF 4 | .895 |  |  |
| Social Influence | SI 1 | .897 | 0.93 | 0.89 |
|  | SI 3 | .949 |  |  |
|  | SI 4 | .890 |  |  |
| Facilitating Conditions | FC 1 | .936 | 0.93 | 0.93 |
|  | FC 2 | .958 |  |  |
|  | FC 3 | .944 |  |  |
|  | FC 5 | .927 |  |  |
| Behavioral Intention | BI 1 | .935 | 0.95 | 0.95 |
|  | BI 2 | .961 |  |  |
|  | BI 3 | .964 |  |  |
|  | BI 4 | .950 |  |  |

## 4.2 The Measurement Model

In order to assess the overall metric model fit, five measures have been applied namely, ratio chi-square to degrees of freedom (X2/d.f.), Root Mean Square of Error Approximation (RMSEA), Comparative Fit Index (CFI), Goodness of Fit Index (GFI), and Adjusted Goodness of Fit Index (AGFI). The final model of this study was obtained through the process including deleting items, since seven teen items (PE 1, PE 5, PE6, EE 1, EE 2, EE 3, EE4, SIN 4, SIN 5, SE 3, SE 4, SE 5, SF 3, SI 2, SI 5, BI 5 and FC 4) have been excluded and re-estimating the model. Consequently, it met all previous goodness of fit measures. Since (X2/d.f.) value is below the 3 threshold [4], RMSEA's value is below the 0.08 threshold [6], GFI value is above the 0.9 threshold [5], AGFI value is above the 0.8 threshold [4], while CFI value is above the 0.9 threshold [20]. Table 3 presents the values of previous model-fit measures.

Table 3 Values of overall model- fit measures

| Model-fit measures index | Recommended values | scores |
| --- | --- | --- |
| Chi-square to degrees of freedom ($X^2$/d.f.) | ≤ 3 | 1.149 |
| Comparative Fit Index (CFI) | ≥ 0.90 | 0.997 |
| Root Mean Square of Error Approximation (RMSEA) | ≤ 0.08 | 0.023 |
| Adjusted Goodness of Fit Index (AGFI) | ≥ 0.80 | 0.912 |
| Goodness of Fit Index (GFI) | ≥ 0.90 | 0.933 |

## 4.3 Structural Model and Results

In the previous section, CFA was performed to assess the model's goodness of fit and loading of the research constructs with items which were used to measure them. In this section, a path analysis for structural model was conducted to examine the hypothesized relationships that help to predict employees' intention to use web based training system. Figure 2 explains the structural model with the assessed path coefficient and the adjusted coefficient of determination (R2) scores, while table 4 shows the overall results of hypotheses' examining.

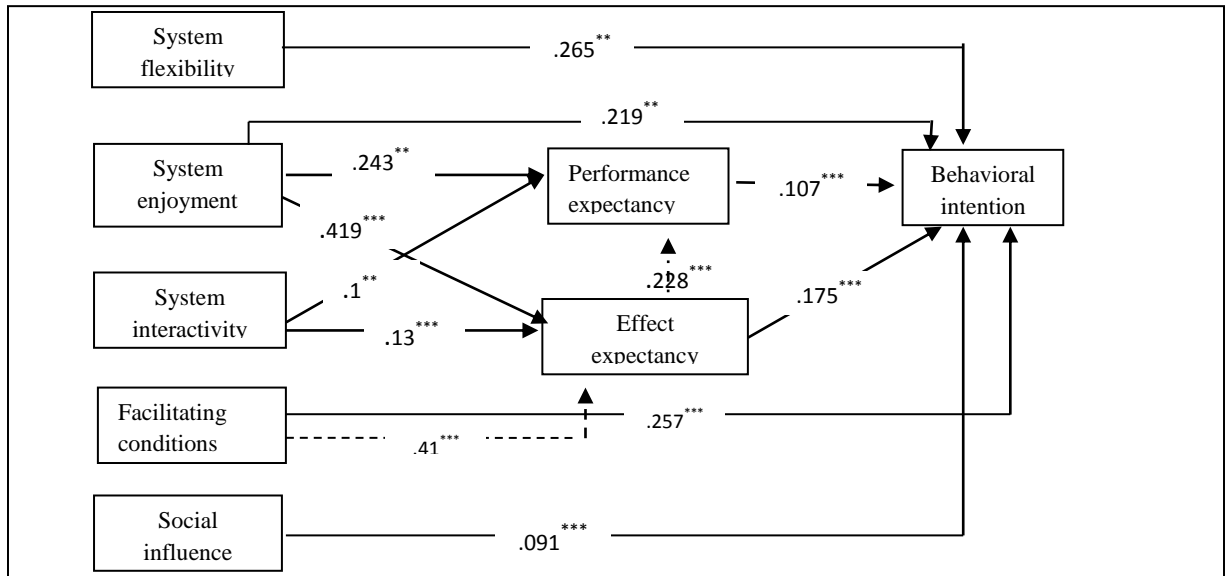

Figure 2. Research structural model

The findings of this study revealed that all of the proposed relationships are accepted and statistically significant. Additionally, according to the modification indices (SEM analysis) there were three new significant relationships: (i) between Effort Expectancy (EE) and Performance Expectance (PE), (ii) between System Flexibility (SF) and Performance Expectance (PE), and (iii) between Facilitating Conditions (FC) and Effort Expectancy (EE). These three relationships have been intervened within the structural model, (see table 4 and figure 1).

Interestingly, the first hypothesis (H1) revealed that performance expectancy will have direct effect on the employees' intention to use web based training system. This hypothesis was accepted, since the statistical result showed that there is strong significant relationship between the performance expectancy and employees' intention to use web based training system (.107***) (Table 4). Additionally, the second relationship (H2) indicated that effort expectancy has direct effect on the employees' intention to use web based training system. This hypothesis was also accepted, since statistical result indicated that there is strong relationship between effort expectancy and employees' intention to use web based training system (.175***) (Table 4). Otherwise, statistical results revealed that there is new significant relationship between effort expectancy and performance expectancy (.228***) (Table 4). This result is consistent with that Taylor and Todd [21] and Davis [10] who indicated that effort expectancy (ease of use) has affected on performance expectancy (usefulness) and user attitude.

Furthermore, third hypothesis (H3) indicated that there is significant relationship between social influence and employees' intention to use web based training system. Consequently, statistical results indicated that there is relationship (0.091***) (Table 4) among social influence and user's intention to use web training (H4). This finding has supported the findings of Venkatesh et al. [24] and Venkatesh and Morris [23]. Focusing in this relationship, it can be assumed that, employees pay much attention about the opinions of other people who are important to them, when they intend to use web training system. Otherwise, the opinions of the people who important for employees (e.g. their managers) influence them to use web training system.

Additionally, fourth hypothesis (H4) indicated that facilitating conditions have direct effect on the employees' intention to use web based training system. This hypothesis was accepted, since the statistical result revealed that there is a strong relationship (.257***) (Table 4). This result has also been confirmed by Thompson et al. [22] and Ajzen [2]. However, it is contrast with Venkatesh et al. [24] who argued that the facilitating conditions does not have effect on an individual's intention to use an information system, but it have direct effect on the actual use beyond that explained by behavioral intention.

Similar to other studies Hsia and Tseng [7]; Nanayakkara [16] and Lim et al. [13], the relationship between Flexibility of web based training and employees' intention to use web training (H5) has been confirmed (.265***) (Table 4). This relation possibly indicates that, a trainer intends to use a web training system, if he/she believes that he/she can access the system from anywhere at any time. In other words, trainees will participate in the e-training process if they believe that they can choose their training equipment and time themselves. Furthermore, similar to Hsia and Tseng [12] study, this study found that there is also relationship between system flexibility and performance expectancy (0.243***) (table 4).

System interactivity is concerned. Since it refers to degree to which employees believe that web based training can provide interactive communication between members of organizations and trainees and between trainees themselves. This study provides evidence that system interactivity has direct effect on the performance expectancy (0.1**) (Table 4) (H6) and effort expectancy (.132***) (Table 4) (H7). That possibly means when the employees intend to use web based training to interact with members of organization (e.g. help disk) and together, they also believe that web based training will enhance their training performance and make the training much easy. This result is in contrast with Abbad et al. [1] and similar to Lim et al. [13] and Davis [10].

As regard to the eighth hypothesis (H8), ninth hypothesis (H9) and tenth hypothesis (H10) which revealed that system enjoyment has a positive impact on perceived performance expectancy, on perceived effort expectancy, and on the employees' intention. These hypotheses were accepted, since statistical results showed that system enjoyment has a strong impact on performance expectancy (.337***) (Table 4); has a strong impact on effort expectancy (.419***) (Table 4); and has direct effect on the employees intention (.219***) (Table 4). These results were supported vary previous studies. Such as Chatzoglou et al. [7] and Abbad et al. [1] found that there are significant relationships between system enjoyment usefulness (performance expectancy), ease of use (effort expectancy) and behavioral intention.

Table 4. hypotheses testing results

| Hypotheses | Path | Path coefficient | remarks |
|---|---|---|---|
| H1 | Performance expectancy and intention | $0.107^{***}$ | accepted |
| H2 | Effort expectancy and intention | $0.175^{***}$ | accepted |
| H3 | Social influence and intention | $0.091^{***}$ | Accepted |
| H4 | Facilitating condition and intention | $0.257^{***}$ | Accepted |
| H5 | System flexibility and intention | $0.265^{***}$ | Accepted |
| H6 | System interactivity and performance expectancy | $0.1^{**}$ | Accepted |
| H7 | System interactivity and effort expectancy | $0.132^{***}$ | Accepted |
| H8 | System enjoyment and performance expectancy | $0.337^{***}$ | Accepted |
| H9 | System enjoyment and effort expectancy | $0.419^{***}$ | Accepted |
| H10 | System enjoyment and intention | $0.219^{***}$ | Accepted |
| **New detected relationships** | | | |
| Effort expectancy and Performance expectancy | | $.228^{***}$ | |
| Facilitating Conditions and Effort Expectancy | | $.410^{***}$ | |
| System Flexibility and Performance Expectancy | | $.243^{***}$ | |

** $P < 0.05$ level, and *** $P < 0.01$ level.

## 5.0 CONCLUSION AND RESEARCH LIMITATIONS

### 5.1 Conclusion

This research has been conducted to collect data from organizations' employees in Jordan; in order to examine the acceptance of web based training system by those employees. The results of this research has indicted that six factors, namely facilitating conditions; performance expectancy, effort expectancy, system flexibility, system enjoyment, and social influence, have direct effect on the employee's intention to use a web based training system. Furthermore, system interactivity, system enjoyment, system flexibility, and facilitating conditions, have affected the performance expectancy and effort expectancy.

These results showed that the employees intend to use web based training system due to improve their training and complete it more quickly, since performance expectancy have strong effect on their intention to use web based training system.

Additionally, one of the study contributions is, this study found that perceived system flexibility impacts intention of the employees to use web based training system, thus, web based training system's designers should assure that the system's components is accessible from anywhere at any time. Further, the employees pay much attention for the opinions of people who are important for them (e.g. their supervisors or their peers); since the result showed that social influence impacts the employees' intention. Knowledge and resources which necessary in the training process are concerned, since the result indicates facilitating conditions have a strong effect on the employees' intention to use web based training system. Therefore, managers should take into their account employees' knowledge and resources which are needed in training process, in order to motivate them and increase their interest to use web based training system.

Furthermore, the employees should feel joyful and can contact other people (e.g. other trainees, trainers or organization's members) during a training process, in order to realize the performance expectancy and effort expectancy of training process, since the statistical result indicate that system enjoyment and system interactivity have direct effect on the performance expectancy and effort expectancy, and have indirect effect on the employees intention to use web based training system.

### 5.2 Research Limitations and Future Research

The first limitation of this research relates to sample size, since small one was taken into consideration (290). Second, other limitation relates to measurement items, whereas just high rate items (α) were taken into consideration. Additionally, further research should pay attention to employees' characteristics (such as, computer anxiety and computer self-efficacy) and assess changing of these characteristics over the time. Furthermore, as prior mention regards lack of relevant studies, thus, more studies should be conducted in this context (acceptance of an information technology by public sector's employees) to support this study's findings.